\documentclass[runningheads]{llncs}
\usepackage{graphicx}
\usepackage{enumitem}
\usepackage{booktabs} 
\usepackage{xspace}
\usepackage[mathscr]{euscript}
\usepackage{algorithm, algorithmicx}
\usepackage[noend]{algpseudocode}
\usepackage{ amssymb }
\usepackage{amsmath}
\usepackage{mathtools}
\usepackage{color}
\usepackage{caption}
\usepackage{subcaption}
\usepackage{microtype}
\usepackage{mdframed}
\usepackage{nicefrac}       
\usepackage{wrapfig}
\usepackage{amsmath,scalerel}
\usepackage{bbm}
\usepackage{relsize}
\usepackage{import}
\usepackage{epsdice}

\newmdtheoremenv[style=2]{mdexample}{Example}
\newmdtheoremenv[style=2]{mdobs}{Observation}
\newmdtheoremenv[style=2]{mddef}{Definition}
\newmdtheoremenv[style=2]{mdtheo}{Theorem}
\newmdtheoremenv[style=2]{mdlem}{Theorem}
\newmdtheoremenv[style=2]{mdcorr}{Corollary}
\newmdtheoremenv[style=2]{mdprop}{Proposition}
\newmdtheoremenv[style=2]{mdprune}{Pruning Rule}
\newmdtheoremenv[style=proofstyle]{mdproof}{Proof}

\DeclareMathAlphabet{\mathpzc}{OT1}{pzc}{m}{it}

\newcommand{\ignore}[1]{}

\newcommand{\mypara}[1]{\paragraph{\bf #1.}}

\newcommand{\Reals}{\mathbb{R}}

\newcommand{\Nat}{\mathbb{N}}

\newcommand{\f}{\varphi}

\newcommand{\eqdef}{\mathrel{\stackrel{\makebox[0pt]{\mbox{\normalfont\tiny def}}}{=}}}

\DeclareMathOperator*{\Prob}{\Pr}

\newcommand{\Bool}{\{0, 1\}}
\newcommand{\BoolCube}{\Bool^n}

\newcommand{\SampleBV}{\overset{ }{\crampedclap{\vx \sim \BoolCube}}}

\newcommand{\KMODELS}{H}
\newcommand{\UNROLL}{U}

\newcommand{\LAST}{\text{last}}

\newcommand{\vx}{\vec{x}}
\newcommand{\vy}{\vec{y}}
\newcommand{\vs}{\vec{s}}
\newcommand{\va}{\vec{a}}
\newcommand{\tausim}[1]{\mathrel{\stackrel{\makebox[0pt]{\mbox{\normalfont\tiny ${#1}$}}}{\sim}}}

\usepackage[colorinlistoftodos]{todonotes}

\setlist{itemsep=1pt, topsep=2pt}

\begin{document}
\title{A Model Counter's Guide to\\Probabilistic Systems}
%
%
\author{
  Marcell Vazquez-Chanlatte\inst{1} \and
  Markus N. Rabe\inst{1} \and
  Sanjit A. Seshia \inst{1}
}

\authorrunning{Vazquez-Chanlatte et al.}
%
\institute{University of California, Berkeley, USA\\
\email{\{marcell.vc, rabe, sseshia\}@eecs.berkeley.edu}\\
}
\maketitle              

\begin{abstract}
In this paper, we systematize the modeling of probabilistic systems for the purpose of analyzing them with model counting techniques.
Starting from unbiased coin flips, we show how to model biased coins, correlated coins, and distributions over finite sets.
From there, we continue with modeling sequential systems, such as Markov chains, and revisit the relationship between weighted and unweighted model counting.
Thereby, this work provides a conceptual framework for deriving \#SAT encodings for probabilistic inference.
\end{abstract}


\section{Introduction}
\vspace{-6px}
Model checking of probabilistic systems, such as Markov Chains, as well as probabilistic inference on
graphical models, capture a diverse range of applications from
biology~\cite{kwiatkowska2008using} to network reliability estimation~\cite{kwiatkowska2003probabilistic}
to learning from unlabeled demonstrations~\cite{vc}.
At their core, such problems often rely on Monte Carlo
methods~\cite{metropolis1949monte}, Belief Propagation~\cite{bishop2006pattern}, and
(explicit-state/BDD-based) probabilistic model
checking~\cite{PRISM,dehnert2017storm,rabe2014symbolic}.
While powerful in specific domains, none of these approaches is a panacea.  
Probabilistic model checking does not scale well to complex systems, and Monte Carlo as well as Belief Propagation requires exponential effort to analyze rare events.
Thus, both methods struggle in settings that involve the analysis of rare events in complex systems.

Model counting is a promising alternative algorithmic approach to analyzing probabilistic models and probabilistic inference.
The recent rapid improvements in SAT-based model counting,
particularly in approximate model
counting~\cite{gomes2007short,chakraborty2013scalable,Soos/2014/Cryptominisat,chakraborty2016algorithmic,achlioptas2018fast,SoosMeel/2019/BIRD},
raise hope that the Herculean improvements in SAT solving could be
leveraged for probabilistic inference. In particular, model counting
may enable us to transition away from Monte Carlo methods and
BDD-based probabilistic model checking, in the same way that SAT
solvers resulted in a transition away from explicit-state/BDD-based
model checking~\cite{biere1999symbolic}.

However, analyzing probabilistic systems with model counting algorithms requires a different approach to modeling than for functional verification.
For example, in SAT-based bounded model checking, each satisfying assignment represents a path of the modeled system.
Since functional verification only requires finding a \emph{single} path, the existence of potentially redundant paths or non-deterministic choices in the encoding is largely irrelevant.
Moreover, because introducing non-determinism and redundant models often simplifies system models, such tricks are frequently employed in system encodings.
By contrast, model counting concerns itself with the model count, i.e. the \emph{number} of satisfying assignments, and hence we must be careful when introducing non-deterministic choices in the model, as they can increase the number of satisfying assignments.


\noindent
\textbf{Contributions:}
In this paper, we systematically develop a framework for modeling
probabilistic systems as model counting problems. The central idea
underlying our framework is that feeding random unbiased coin flips
into a Boolean predicates simulates a biased coin flip. Based on this
deceptively simple observation, the rest of the paper develops
numerous gadgets, which, when combined, can encode arbitrary
distributions over finite sets, as well as Markov Chains.
\begin{enumerate}
  \itemsep0em
\item A framework for deriving unweighted model counting encodings for
  queries about probabilistic systems such as Markov Chains. Two
  key features of this framework are (i) the focus on Boolean functions rather
  than constraints. This alternative focus facilitates composition and
  preserves model counts. (ii) the use of sequential circuits,
  in particular the decomposition given in Fig.~\ref{fig:monitor},
  for encoding Markov Chains.
\item An alternative perspective on the reduction from weighted to
  unweighted model counting provided by~\cite{weighted2unweighted},
  which makes apparent that the central gadget used is the less-than
  operator on unsigned integers.
\item Three algorithms for encoding queries about distributions over finite
  sets as model counting problems. The first builds on the weighted to
  unweighted reduction given in~\cite{weighted2unweighted} while the
  other two illustrate how our framework facilitates interfacing with the
  larger random number generation literature.
\end{enumerate}

The rest of this paper is organized as follows: We begin by
establishing the connection between model counting and unbiased coin-flips.
Then, in Section~\ref{sec:rand_inputs}, we demonstrate how to use unbiased coins to model biased coins and correlated coins.  
In Section~\ref{sec:seq_circuits}, we show how to use these components to model sequential probabilistic systems, such as Markov Chains.
We demonstrate how to model arbitrary distributions over finite sets in Section~\ref{sec:arb_dist}, including the binomial distribution, which is fundamental to probabilistic inference.
Finally, we revisit the relationship of weighted and unweighted model counting in Section~\ref{sec:rel_unweighted}.

\vspace{-10px}
\section{Related Work}
\vspace{-10px}
This work connects with literature in two primary ways. 
First and foremost, we provide an encoding of probabilistic systems as sequential circuits, which when unrolled, are suited to be analyzed with model counting algorithms.
Numerous encodings of probabilistic systems as \emph{weighted} model counting problems have been proposed~\cite{sang2005performing,chavira2008probabilistic,dalvi2007efficient}, which has then spurred the adaptation of a number of unweighted model counting algorithms to solve weighted model counting problems~\cite{sang2005performing,choi2013dynamic,chakraborty2014distribution}.
Unfortunately, such adaptations require expert knowledge of the inner workings of model counters, making the transfer of advancements from unweighted model counting to weighted model counting difficult.
Hence, techniques for efficient and automated reductions from weighted model counting to unweighted model counting have been proposed~\cite{weighted2unweighted}.
In many ways, this article continues in this direction, by providing a framework for encoding probabilistic systems and inferences on said systems \emph{directly} into unweighted model counting problems.
Such reductions are particularly appealing given that, to our knowledge, there is no major algorithmic advantage in using weighted over unweighted model counting algorithms.

Further, this work is intimately related to the work on simulating discrete distributions using a stream of random bits.
This framework,
called the random bit model and first introduced by von
Neumann~\cite{von1963various}, has gone on to spawn numerous
techniques (for a more detailed survey, we point the reader to~\cite{lumbroso2013optimal}).
One of the goals in this work has been to illustrate how to draw
from this vast literature to create new encodings of probabilistic
circuits, e.g., in Section~\ref{sec:arb_dist}, we illustrate
how to a binomial distribution as well as encode Knuth and Yao's~\cite{knuthyaodice} classic algorithm
into a sequential circuit.


\vspace{-10px}
\section{Circuits, Coin Flips, and Model Counting}\label{sec:rand_inputs}
\vspace{-10px}
In the sequel, we develop a framework for analyzing probabilistic systems via model counting.
We begin by defining bit-vectors and bit-vector predicates.

\begin{figure}[h]
  \begin{subfigure}[b]{0.45\linewidth}
    
    \centering \def\svgwidth{0.7in} 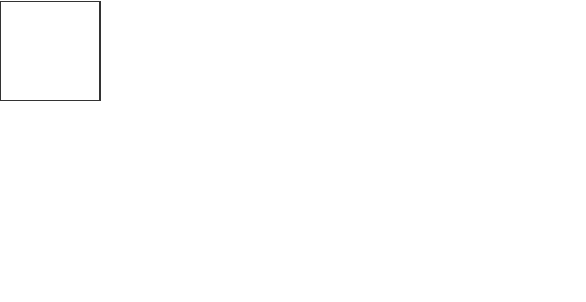
    \caption{Example concatenation of $3$ and $2$ length bit-vectors to form a $5$-bit-vector.}
    \label{fig:bitvector}
  \end{subfigure}
  \hfill
  \begin{subfigure}[b]{0.45\linewidth}
    \centering
    \scalebox{0.7}{
      \def\svgwidth{0.5in} 
\begingroup%
  \makeatletter%
  \providecommand\color[2][]{%
    \errmessage{(Inkscape) Color is used for the text in Inkscape, but the package 'color.sty' is not loaded}%
    \renewcommand\color[2][]{}%
  }%
  \providecommand\transparent[1]{%
    \errmessage{(Inkscape) Transparency is used (non-zero) for the text in Inkscape, but the package 'transparent.sty' is not loaded}%
    \renewcommand\transparent[1]{}%
  }%
  \providecommand\rotatebox[2]{#2}%
  \newcommand*\fsize{\dimexpr\f@size pt\relax}%
  \newcommand*\lineheight[1]{\fontsize{\fsize}{#1\fsize}\selectfont}%
  \ifx\svgwidth\undefined%
    \setlength{\unitlength}{124.50117937bp}%
    \ifx\svgscale\undefined%
      \relax%
    \else%
      \setlength{\unitlength}{\unitlength * \real{\svgscale}}%
    \fi%
  \else%
    \setlength{\unitlength}{\svgwidth}%
  \fi%
  \global\let\svgwidth\undefined%
  \global\let\svgscale\undefined%
  \makeatother%
  \begin{picture}(1,1.10971424)%
    \lineheight{1}%
    \setlength\tabcolsep{0pt}%
    \put(0,0){\includegraphics[width=\unitlength,page=1]{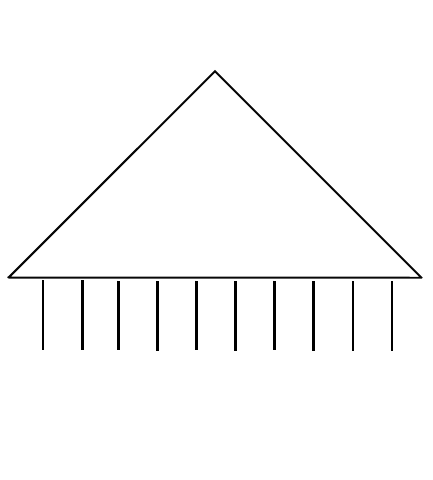}}%
    \put(0.44277151,0.64113007){\color[rgb]{0,0,0}\makebox(0,0)[lt]{\lineheight{1.25}\smash{\begin{tabular}[t]{l}$\f$\end{tabular}}}}%
    \put(0,0){\includegraphics[width=\unitlength,page=2]{imgs/circ.pdf}}%
    \put(0.40103364,0.00000002){\color[rgb]{0,0,0}\makebox(0,0)[lt]{\lineheight{1.25}\smash{\begin{tabular}[t]{l}n\end{tabular}}}}%
  \end{picture}%
\endgroup%

    }
    \caption{A n-ary bit-vector predicate as a circuit with $n$ inputs.}
    \label{fig:bool_pred}
  \end{subfigure}
  \caption{Illustrations to accompany Definitions~\ref{def:bv} and~\ref{def:bv_pred}.}
\end{figure}

\begin{mddef}\label{def:bv}
  A \textbf{$n-$bit-vector}, $\vx \in \BoolCube$, is a tuple of
  $n\in\Nat$ Boolean values. The \textbf{concatenation} of a
  $n$-bit-vector, $\vx$, and a $m$-bit-vector, $\vx'$, is an
  $(n+m)$-bit-vector, $\vx.\vx'$, where the first $n$ bits form $\vx$ and
  the final $m$ bits form $\vx'$ (see Fig.~\ref{fig:bitvector}).
\end{mddef}

To avoid clutter, if $n=m=1$, we simply write $xx'$. Thus,
we denote decomposing an $n-$bit-vector, $\vx$, individual bits via $\vx =
x_1x_2\ldots x_n$.
Next, we define a model counter's main object of study, predicates over bit-vectors.

\begin{mddef}\label{def:bv_pred}
  A \textbf{n-ary bit-vector predicate} maps $n$-bit vectors to\\
  $\Bool \subseteq \Reals$, e.g,
  \begin{equation}\label{eq:bool_pred}
    \f : \BoolCube \to \Bool.
  \end{equation}
  If $\f(\vx) = 1$, we additionally call $\vx$ a \textbf{model} of
  $\f$.  We define the \textbf{model count} of $\f$, denoted $\#(\f)$,
  as the number of models of $\f$, i.e,
  \begin{equation}\label{eq:count}
    \#(\f) \eqdef \sum_{\vx\in \BoolCube} \f(\vx).
  \end{equation}
\end{mddef}

  When the number of inputs of a predicate $\f$ is
  unambiguous, we shall write $\f(\vx)$ as a logical sentence over
  $\vx$, where True is mapped to 1 and False is mapped to 0.

\begin{example}
  Let $\f : \Bool^{10} \to \Bool$ denote the $10$ input map,
  \begin{equation}
    \f(\vx) = x_1 \wedge x_7 \eqdef
    \begin{cases}
      1 & \text{if } x_1 \wedge x_7\\
      0 & \text{otherwise}
    \end{cases}.
  \end{equation}
  Thus $\f(\vx) = 1$ iff $x_1$ and $x_7$ are
  True (i.e., $1$). Further observing that there are $8$ other ``don't care'' inputs, each with
  two possible values, yields $\#(\f) = 2^8$.
\end{example}

\begin{example}\label{ex:lt_models}
  Given $n \in \Nat$, let $k$ be an integer between $0$ and $2^n - 1$ and let
  $\f : \Bool^n \to \Bool$ denote,
  \begin{equation}
    \f(\vx) = \vx < k,
  \end{equation}
  where $\vx$ is interpreted as an integer between $0$ and
  $2^n - 1$. Observe that $\#(\f) = k$ since there are only
  $k$ unsigned integers less than $k$.
\end{example}
\vspace{-4px}
Next, observe that a circuit can be made probabilistic by feeding the
results of random coin flips as inputs.  To this end, we introduce
notation for the process of generating a bit-vector using $n$ unbiased
coin flips.
\begin{mddef}
  Denote by $x_1x_2\ldots x_n \sim \BoolCube$ the act of creating an
  $n-$bit-vector by flipping $n$ independent unbiased coins with
  \begin{equation}
    \label{eq:random_bit_flip}
    \Prob_{\SampleBV}(x_i = 0) = \Prob_{\SampleBV}\; (x_i = 1) = \frac{1}{2},
  \end{equation}
  and thus, the probability of drawing any particular bit-vector,
  $\vx^*$ is:
  \begin{equation}
    \label{eq:random_draw}
    \Prob_{\SampleBV}(\vx = \vx^*) = \frac{1}{2^n}.
  \end{equation}
\end{mddef}
Our framework for studying random inputs to bit-vector functions
relies on the following key (though unsurprising) observation.
\begin{mdobs}\label{obs:1}
  Given an $n$-ary bit-vector predicate, $\f$, if one flips $n$
  independent unbiased coins, $\vx \sim \BoolCube$,
  the probability that $\f(\vx) = 1$ is equal
  to the fraction of $n$-bit-vectors that are models of $\f$, i.e,
  
  \begin{equation}\label{eq:model_count_coins}
    \Prob_{\SampleBV}\;(\f(\vx) = 1)
    = \sum_{\vx \in \BoolCube}\frac{1}{2^n}\f(\vx)\\
    = \frac{\#(\f)}{2^n}.
  \end{equation}  
\end{mdobs}

Therefore, if one wishes to compute $\displaystyle\Prob_{\SampleBV}(\f(\vx) = 1)$ for some complicated $\f$, it suffices to use a
model counter to compute (or approximate) $\#(\f)$.  While
straightforward, the power of this observation is only truly realized
when one starts composing bit-vector predicates and reusing inputs. We
illustrate this through a series of observations.
\begin{mdobs}\label{obs:2}
  Using \eqref{eq:model_count_coins}, $\f$ can be reinterpreted as a process
  to turn $n$ unbiased coins into a \emph{biased} coin.
\end{mdobs}
To emphasize Observation~\ref{obs:2}, we shall denote by $x \sim \f$
the process of drawing a biased coin, $x \in \Bool$, using the
distribution given in~\eqref{eq:model_count_coins}.

\begin{figure}[t]
  \begin{subfigure}[b]{0.45\linewidth}
    \centering \def\svgwidth{0.75in} 
\begingroup%
  \makeatletter%
  \providecommand\color[2][]{%
    \errmessage{(Inkscape) Color is used for the text in Inkscape, but the package 'color.sty' is not loaded}%
    \renewcommand\color[2][]{}%
  }%
  \providecommand\transparent[1]{%
    \errmessage{(Inkscape) Transparency is used (non-zero) for the text in Inkscape, but the package 'transparent.sty' is not loaded}%
    \renewcommand\transparent[1]{}%
  }%
  \providecommand\rotatebox[2]{#2}%
  \newcommand*\fsize{\dimexpr\f@size pt\relax}%
  \newcommand*\lineheight[1]{\fontsize{\fsize}{#1\fsize}\selectfont}%
  \ifx\svgwidth\undefined%
    \setlength{\unitlength}{164.72959864bp}%
    \ifx\svgscale\undefined%
      \relax%
    \else%
      \setlength{\unitlength}{\unitlength * \real{\svgscale}}%
    \fi%
  \else%
    \setlength{\unitlength}{\svgwidth}%
  \fi%
  \global\let\svgwidth\undefined%
  \global\let\svgscale\undefined%
  \makeatother%
  \begin{picture}(1,0.61366806)%
    \lineheight{1}%
    \setlength\tabcolsep{0pt}%
    \put(0,0){\includegraphics[width=\unitlength,page=1]{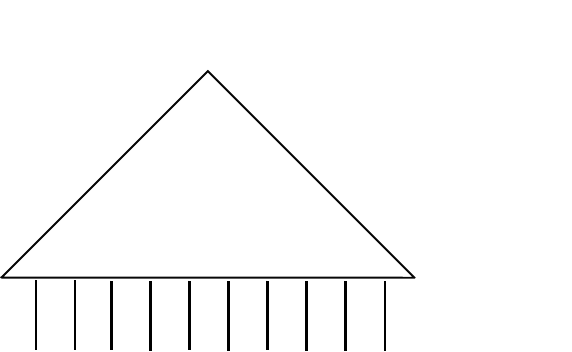}}%
    \put(0.24033224,0.25951626){\color[rgb]{0,0,0}\makebox(0,0)[lt]{\lineheight{1.25}\smash{\begin{tabular}[t]{l}$\f$\end{tabular}}}}%
    \put(0,0){\includegraphics[width=\unitlength,page=2]{imgs/circ2.pdf}}%
    \put(0.60456552,0.25951626){\color[rgb]{0,0,0}\makebox(0,0)[lt]{\lineheight{1.25}\smash{\begin{tabular}[t]{l}$\f'$\end{tabular}}}}%
    \put(0,0){\includegraphics[width=\unitlength,page=3]{imgs/circ2.pdf}}%
  \end{picture}%
\endgroup%

    \caption{By sharing inputs, two bit-vector predicates, which model  biased coins, can be used
      to model a pair of \emph{correlated} coin flips.}
    \label{fig:bool_pred}
  \end{subfigure}
  \hfill
  \begin{subfigure}[b]{0.45\linewidth}
    \centering \def\svgwidth{0.6in} 
\begingroup%
  \makeatletter%
  \providecommand\color[2][]{%
    \errmessage{(Inkscape) Color is used for the text in Inkscape, but the package 'color.sty' is not loaded}%
    \renewcommand\color[2][]{}%
  }%
  \providecommand\transparent[1]{%
    \errmessage{(Inkscape) Transparency is used (non-zero) for the text in Inkscape, but the package 'transparent.sty' is not loaded}%
    \renewcommand\transparent[1]{}%
  }%
  \providecommand\rotatebox[2]{#2}%
  \newcommand*\fsize{\dimexpr\f@size pt\relax}%
  \newcommand*\lineheight[1]{\fontsize{\fsize}{#1\fsize}\selectfont}%
  \ifx\svgwidth\undefined%
    \setlength{\unitlength}{196.00634477bp}%
    \ifx\svgscale\undefined%
      \relax%
    \else%
      \setlength{\unitlength}{\unitlength * \real{\svgscale}}%
    \fi%
  \else%
    \setlength{\unitlength}{\svgwidth}%
  \fi%
  \global\let\svgwidth\undefined%
  \global\let\svgscale\undefined%
  \makeatother%
  \begin{picture}(1,0.92517053)%
    \lineheight{1}%
    \setlength\tabcolsep{0pt}%
    \put(0,0){\includegraphics[width=\unitlength,page=1]{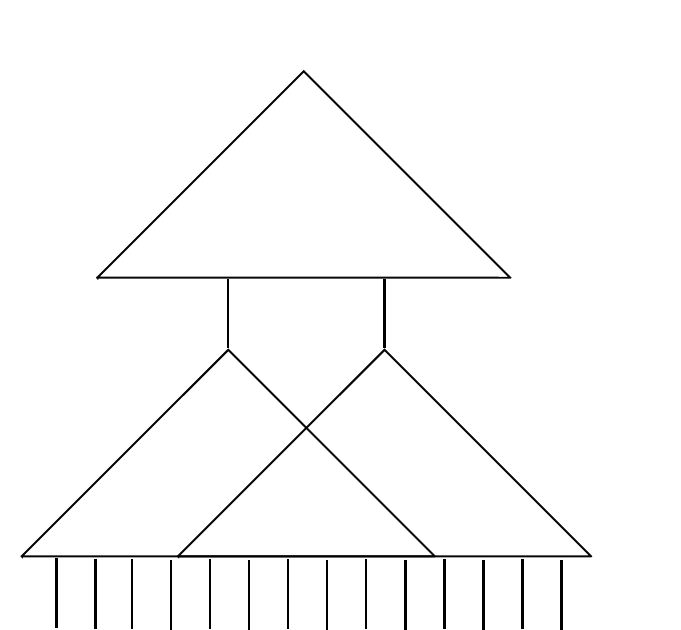}}%
    \put(0.41306287,0.59409818){\color[rgb]{0,0,0}\makebox(0,0)[lt]{\lineheight{1.25}\smash{\begin{tabular}[t]{l}$\f''$\end{tabular}}}}%
    \put(0,0){\includegraphics[width=\unitlength,page=2]{imgs/circ3.pdf}}%
    \put(0.92085621,0.18972925){\color[rgb]{0,0,0}\makebox(0,0)[lt]{\lineheight{1.25}\smash{\begin{tabular}[t]{l}F\end{tabular}}}}%
  \end{picture}%
\endgroup%

    \caption{Feeding correlated biased coin flips into a bit-vector
      predicate yields a new bit-vector predicate, and thus models a
      biased coin flip.}
    \label{fig:bool_pred}
  \end{subfigure}
  \caption{Illustrations of Observations~\ref{obs:2} and~\ref{obs:3}.}
\end{figure}

\begin{mdobs}\label{obs:3}
  If the results of some coin flips are shared, $F: \vx \mapsto (\f(\vx), \f'(\vx))$, then
  $F : \Bool^n \to \Bool^2$ models \emph{correlated} coin flips.
\end{mdobs}

As above, inspired by Observation~\ref{obs:3}, given a map between bit-vectors,
$F : \Bool^n \to \Bool^m$, we denote by $\vx \sim F$ the process of drawing $m$
correlated biased coin flips. In particular, if $\f_i(\vx) = F(\vx)_i$, then
$\vx$ is the concatenation of $m$ bit-vectors such that, $x_i \sim \f_i$.
Together, Observations~\ref{obs:2} and~\ref{obs:3} enable
studying complex distributions via model counting.

\begin{example}\label{ex:mux}
  Let $\f$ and $\f'$ denote the following $3$-bit bit-vector
  predicates,
  \begin{equation}\label{eq:range_mux}
    \begin{split}
      &\f : \Bool^3 \to \Bool\\
      &\f(\vx) \eqdef \vx = 3
    \end{split}
    ~~~~~~~~~~~~    
    \begin{split}
      &\f' : \Bool^3 \to \Bool\\
      &\f'(\vx) \eqdef \vx > 3
    \end{split},
  \end{equation}
  where $\vx$ is interpreted as an unsigned integer.

  Next, define $\f \times \f'$ as the product of $\f$ and $\f'$, i.e.,
  $[\f \times \f'] : \vx \mapsto \Big(\f(\vx), \f'(\vx)\Big)$. The resulting
  map is illustrated in Fig.~\ref{fig:range_mux}.
  \begin{figure}[t]
    \centering
    \scalebox{0.7}{
      \def\svgwidth{2.6in} 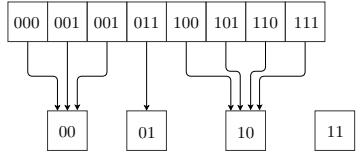
    }
    \caption{Visualization of $[\f \times \f'](\vx)$. The bit sequences
      have the most significant bit on the left and the least
      significant bit on the right, e.g., $011 = 3$.}
    \label{fig:range_mux}
  \end{figure}

  Again, note that because $\f$ and $\f'$ share inputs, then the
  biased coins they model are correlated (Observation~\ref{obs:3}). In
  particular, using Fig.~\ref{fig:range_mux}, we see that
  $\f \times \f'$ induces the following distribution over $2$-bit-vectors:
  \begin{equation}\label{eq:range_mux_bv}
    \Prob_{\mathclap{\vy \sim \f\times \f'}}\big(\vy = k\big) =
    \begin{cases}
      3/8 & \text{if } k = 0\\
      1/8 & \text{if } k = 1\\
      1/2 & \text{if } k = 2\\
      0 & \text{otherwise}.
    \end{cases}
  \end{equation}

  Now suppose one wishes to compute the probability that $k > 0$
  under~\eqref{eq:range_mux_bv}.  By~\eqref{eq:model_count_coins}, it
  suffices to compute the model count of $[\vx > 0]$ composed with
  $\f\times \f'$,
  \begin{equation} \label{eq:8}
    \Prob_{\mathclap{\vy \sim \f\times \f'}}\big(\vy > 0\big)
    = \frac{\#\Big([\vx > 0]\circ[\f\times
      \f']\Big)}{2^3}.
  \end{equation}
  Of course, in this case, it is easy to look at
  Fig.~\ref{fig:range_mux} to determine that\\
  $\#\Big([\vx > 0]\circ[\f\times \f']\Big) = 5$. However, in general,
  with bigger circuits and more complicated properties, this explicit
  reduction to model counting proves incredibly useful.
\end{example}

The framework developed so far has focused on modeling probability
distributions where the probability masses are (integer) multiples of
$\frac{1}{2^n}$. Of course, many examples violate this assumption,
e.g, a coin with a $\frac{1}{3}$ bias towards heads. To handle such
distributions, we adapt our framework to condition on certain coin
flip outcomes not occurring. Note that via the chain rule, any
predicate over an input conditioned distribution can be studied using
two model counting queries.
\begin{mdframed}
\begin{proposition}
  Let $\f : \Bool^n \to \Bool$ and $\psi : \Bool^n \to \Bool$ denote
  any two bit-vector predicates. Then,
  \begin{equation}\label{eq:model_count_conditioned}
    \Prob_{\SampleBV}(\f(\vx)=1~|~\psi(\vx)=1) = \frac{\#(\f \wedge \psi)}{\#(\psi)}.
  \end{equation}
\end{proposition}
\end{mdframed}

\begin{proof}
  By the chain rule,
  \begin{equation}
    \label{eq:chain} \Prob_{\SampleBV}(\f(\vx)=1~|~\psi(\vx)=1)\cdot
\Prob_{\SampleBV}(\psi(\vx)=1) = \Prob_{\SampleBV}([\f \wedge \psi](\vx) =
1).
  \end{equation}
  Replacing the unconditioned probabilities using~\eqref{eq:model_count_coins} gives,
  \begin{equation}
    \Prob_{\SampleBV}(\f(\vx)=1~|~\psi(\vx)=1)\frac{\#(\psi)}{2^n} = \frac{\#(\f\wedge \psi)}{2^n}.
  \end{equation}
  Multiplying both sides by $2^n$ and rearranging yields~\eqref{eq:model_count_conditioned}. \qed
\end{proof}

To avoid notational clutter, we shall frequently write
\eqref{eq:model_count_conditioned} using the sampling notation, $y \sim \f$, previously
introduced, but additionally condition on $\psi$,
\begin{mdframed}
\begin{equation}
  \Prob_{y \sim \f}(y~|~\psi)\eqdef\Prob_{\vx \sim \BoolCube}\big(\f(\vx)=1~|~\psi(\vx)=1\big)
\end{equation}  
\end{mdframed}
\begin{example}\label{ex:dice}
  Again, suppose we seek to find a pair $\f, \psi$ that encodes a
  biased coin with probability $\nicefrac{1}{3}$ of coming up $1$.
  Observe that this can be accomplished by letting
  $\f(\vx) \eqdef (\vx = 0), \psi(\vx) \eqdef (\vx < 3)$, such that,
  \begin{equation}
    \Prob_{y \sim \f}(y_i~|~\psi) = \frac{\#(\vx = 0 \wedge \vx < 3)}{\#(\vx  < 3)} = \frac{1}{3},
  \end{equation}
  where $\vx \in \Bool^2$ is encoded as an unsigned integer.
\end{example}
Note that in many contexts, $\#(\psi)$ can be precomputed, sometimes
even without the use of a model counting algorithm.

\mypara{Encoding Rational Coins} Example~\ref{ex:dice} can be generalized
to encode an arbitrary coin with a rational bias. Namely,
consider a coin, $y$, such that $\Pr(y = 1) =\frac{k}{m}$,
for some $k, m \in \Nat$. Letting $n$
be the smallest integer such that $m \leq 2^n$, and recalling that
$\vx < k$ has exactly $k$ models (Ex.~\ref{ex:lt_models}), observe
that $y$ corresponds to feeding $n$ unbiased coins into $\f(\vx) \eqdef \vx < k$
and conditioning on $\psi(\vx) \eqdef \vx < m$. Finally, observing that
$\vx < k$ implies that $\vx < m$ yields,
\begin{equation}\label{eq:arb_coin}
  \Prob_{y \sim \f}(y = 1~|~\psi) = \frac{\#(\vx < k)}{\#(\vx < m)} = \frac{k}{m}.
\end{equation}
Finally, observe since that $\Prob_{\vy \sim F}( \f(\vy)~|~\psi, \psi') = \Prob_{\vy \sim F}(\f(\vy) = 1~|~\psi \wedge \psi')$, Eq.~\eqref{eq:arb_coin} naturally extends to
modeling multiple input conditioned coin flips.
Of course, biased coins are not very interesting by
themselves. Nevertheless, as illustrated in Ex.~\ref{ex:mux}, feeding
multiple correlated coin flips into another circuit enables
studying more sophisticated objects. Further, as the next
section illustrates, by incorporating a notion of state, the framework
developed above enables answering non-trivial queries about
probabilistic systems via model counting.
  

\section{Sequential Circuits}\label{sec:seq_circuits}

Ultimately, we want study sequential probabilistic systems, such as Markov Chains, probabilistic regular languages, and random walks.
While the processes we studied in the previous section involved only an a-priori fixed number of coin flips, sequential systems in general may consume an arbitrary number of bits.
We can thus not anymore rely on Boolean predicates, but need to extend our framework.
In the following, we thus introduce sequential circuits and show how to employ them for modeling sequential probabilistic systems.


\begin{figure}[h]
  \begin{subfigure}[t]{0.45\linewidth}
    \centering \def\svgwidth{0.9in} 
\begingroup%
  \makeatletter%
  \providecommand\color[2][]{%
    \errmessage{(Inkscape) Color is used for the text in Inkscape, but the package 'color.sty' is not loaded}%
    \renewcommand\color[2][]{}%
  }%
  \providecommand\transparent[1]{%
    \errmessage{(Inkscape) Transparency is used (non-zero) for the text in Inkscape, but the package 'transparent.sty' is not loaded}%
    \renewcommand\transparent[1]{}%
  }%
  \providecommand\rotatebox[2]{#2}%
  \newcommand*\fsize{\dimexpr\f@size pt\relax}%
  \newcommand*\lineheight[1]{\fontsize{\fsize}{#1\fsize}\selectfont}%
  \ifx\svgwidth\undefined%
    \setlength{\unitlength}{209.72960945bp}%
    \ifx\svgscale\undefined%
      \relax%
    \else%
      \setlength{\unitlength}{\unitlength * \real{\svgscale}}%
    \fi%
  \else%
    \setlength{\unitlength}{\svgwidth}%
  \fi%
  \global\let\svgwidth\undefined%
  \global\let\svgscale\undefined%
  \makeatother%
  \begin{picture}(1,0.89748651)%
    \lineheight{1}%
    \setlength\tabcolsep{0pt}%
    \put(0,0){\includegraphics[width=\unitlength,page=1]{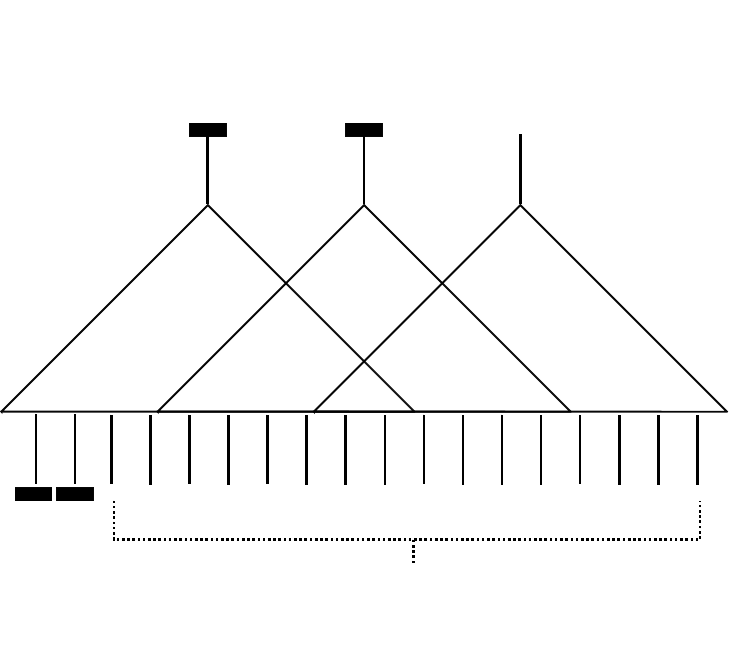}}%
    \put(0.54662482,0.04040122){\color[rgb]{0,0,0}\makebox(0,0)[lt]{\lineheight{1.25}\smash{\begin{tabular}[t]{l}n\end{tabular}}}}%
    \put(0,0){\includegraphics[width=\unitlength,page=2]{imgs/seq_circ.pdf}}%
    \put(0.04754179,0.04405785){\color[rgb]{0,0,0}\makebox(0,0)[lt]{\lineheight{1.25}\smash{\begin{tabular}[t]{l}p\end{tabular}}}}%
    \put(0,0){\includegraphics[width=\unitlength,page=3]{imgs/seq_circ.pdf}}%
    \put(0.67871167,0.80932889){\color[rgb]{0,0,0}\makebox(0,0)[lt]{\lineheight{1.25}\smash{\begin{tabular}[t]{l}m\end{tabular}}}}%
    \put(0,0){\includegraphics[width=\unitlength,page=4]{imgs/seq_circ.pdf}}%
    \put(0.36759676,0.81737499){\color[rgb]{0,0,0}\makebox(0,0)[lt]{\lineheight{1.25}\smash{\begin{tabular}[t]{l}p\end{tabular}}}}%
  \end{picture}%
\endgroup%

    \caption{A sequential circuit where the first $p$ bits of input
      and output are marked with black rectangles to indicate that
      they represent the previous and next state, respectively.}
    \label{fig:seq_circ}
  \end{subfigure}
  \hfill
  \begin{subfigure}[t]{0.45\linewidth}
    \centering \def\svgwidth{1.5in} 
\begingroup%
  \makeatletter%
  \providecommand\color[2][]{%
    \errmessage{(Inkscape) Color is used for the text in Inkscape, but the package 'color.sty' is not loaded}%
    \renewcommand\color[2][]{}%
  }%
  \providecommand\transparent[1]{%
    \errmessage{(Inkscape) Transparency is used (non-zero) for the text in Inkscape, but the package 'transparent.sty' is not loaded}%
    \renewcommand\transparent[1]{}%
  }%
  \providecommand\rotatebox[2]{#2}%
  \newcommand*\fsize{\dimexpr\f@size pt\relax}%
  \newcommand*\lineheight[1]{\fontsize{\fsize}{#1\fsize}\selectfont}%
  \ifx\svgwidth\undefined%
    \setlength{\unitlength}{123.32923222bp}%
    \ifx\svgscale\undefined%
      \relax%
    \else%
      \setlength{\unitlength}{\unitlength * \real{\svgscale}}%
    \fi%
  \else%
    \setlength{\unitlength}{\svgwidth}%
  \fi%
  \global\let\svgwidth\undefined%
  \global\let\svgscale\undefined%
  \makeatother%
  \begin{picture}(1,0.53198307)%
    \lineheight{1}%
    \setlength\tabcolsep{0pt}%
    \put(0,0){\includegraphics[width=\unitlength,page=1]{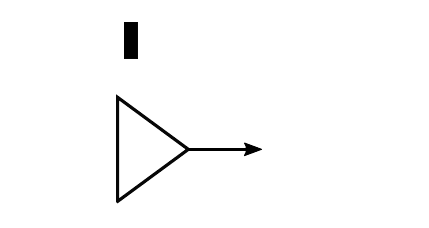}}%
    \put(-0.00233591,0.11124983){\color[rgb]{0,0,0}\makebox(0,0)[lt]{\lineheight{1.25}\smash{\begin{tabular}[t]{l}x\end{tabular}}}}%
    \put(0,0){\includegraphics[width=\unitlength,page=2]{imgs/historically.pdf}}%
    \put(0.48815869,0.09090045){\color[rgb]{0,0,0}\makebox(0,0)[lt]{\lineheight{1.25}\smash{\begin{tabular}[t]{l}historically x\end{tabular}}}}%
    \put(0.30701818,0.16024779){\color[rgb]{0,0,0}\makebox(0,0)[lt]{\lineheight{1.25}\smash{\begin{tabular}[t]{l}$\wedge$\end{tabular}}}}%
    \put(0.32344316,0.48469477){\color[rgb]{0,0,0}\makebox(0,0)[lt]{\lineheight{1.25}\smash{\begin{tabular}[t]{l}1\end{tabular}}}}%
  \end{picture}%
\endgroup%

    \caption{Sequential circuit testing if $x$ is currently $1$
      \emph{and} ($\wedge$) if $x$ has historically been $1$.  In this
      illustration, the latch ($\vs_0 = 1$) is shown cutting the
      cyclic dependency of the circuit.}
    \label{fig:histx}
  \end{subfigure}
  \caption{Sequential Circuit illustration and example.}
\end{figure}

\begin{mddef}
  Let $n, m,$ and $p$ denote natural numbers.  A \textbf{sequential
    circuit} is a tuple, $C = (\vs_0, F)$, where $F : \Bool^{p + n} \to \Bool^{p + m}$ is the transition function, and
  $\vs_0 \in \Bool^p$ is the initial state.

  \noindent
  Further, to every sequence of inputs $\va_1, \va_2, \ldots \in \Bool^n$, we associate a sequence
  of states $\vs_1, \vs_2, \ldots \in \Bool^p$ and outputs $\vy_1, \vy_2, \ldots \in \Bool^m$ by:
  \begin{equation}
    \vs_{i}.\vy_i = F(\vs_{i-1}, \va_i)
  \end{equation}
  Finally, if $m = 1$, we refer to $C$ as a \textbf{monitor}.
\end{mddef}
\begin{example}
  Figure~\ref{fig:histx} illustrates a sequential circuit that checks if
  $x$ has been constantly $1$. Formally if $\f(\vx) \eqdef x_0 \wedge
  x_1$, then $F(\vx) \eqdef \f(\vx).\f(\vx)$ and $\vs_0 = 1$. Note that
  as circuits can reuse outputs, in Fig.~\ref{fig:histx} $\f(\vx)$ is only
  computed once.
\end{example}

Now observe that we can reduce the execution of a fixed number of steps of a sequential circuit, $C = (\vs_0, F)$, back to a bit-vector function simply by composing $F$ with itself, akin to bounded model checking~\cite{biere1999symbolic}.

\begin{figure}[h]
  \centering
  \def\svgwidth{3.2in}
\begingroup%
  \makeatletter%
  \providecommand\color[2][]{%
    \errmessage{(Inkscape) Color is used for the text in Inkscape, but the package 'color.sty' is not loaded}%
    \renewcommand\color[2][]{}%
  }%
  \providecommand\transparent[1]{%
    \errmessage{(Inkscape) Transparency is used (non-zero) for the text in Inkscape, but the package 'transparent.sty' is not loaded}%
    \renewcommand\transparent[1]{}%
  }%
  \providecommand\rotatebox[2]{#2}%
  \newcommand*\fsize{\dimexpr\f@size pt\relax}%
  \newcommand*\lineheight[1]{\fontsize{\fsize}{#1\fsize}\selectfont}%
  \ifx\svgwidth\undefined%
    \setlength{\unitlength}{1091.49031511bp}%
    \ifx\svgscale\undefined%
      \relax%
    \else%
      \setlength{\unitlength}{\unitlength * \real{\svgscale}}%
    \fi%
  \else%
    \setlength{\unitlength}{\svgwidth}%
  \fi%
  \global\let\svgwidth\undefined%
  \global\let\svgscale\undefined%
  \makeatother%
  \begin{picture}(1,0.28445172)%
    \lineheight{1}%
    \setlength\tabcolsep{0pt}%
    \put(0,0){\includegraphics[width=\unitlength,page=1]{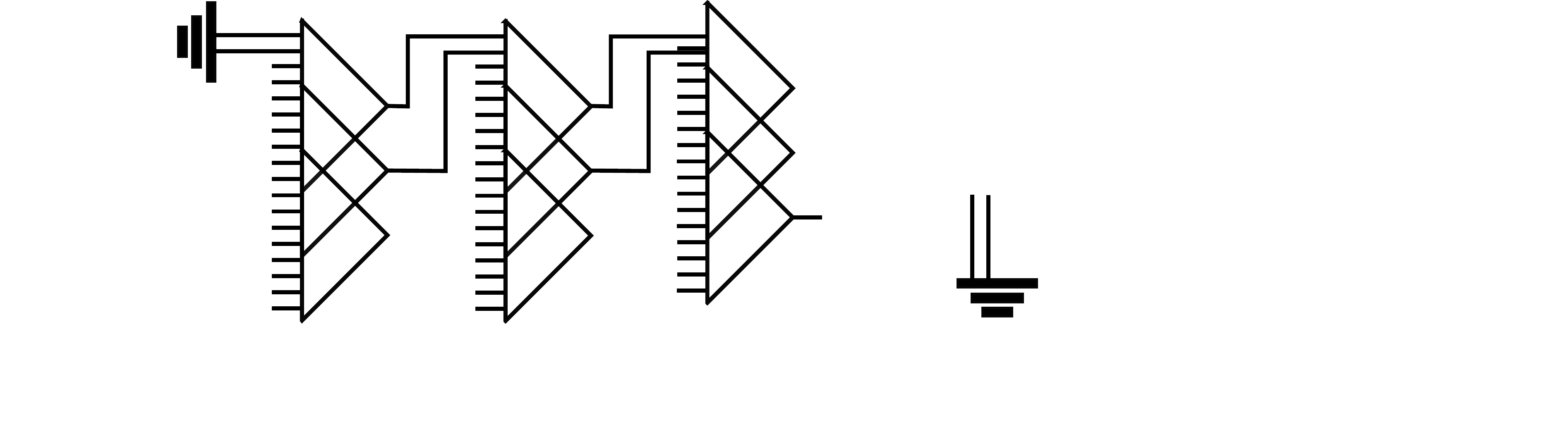}}%
    \put(0.55830314,0.15880785){\color[rgb]{0,0,0}\makebox(0,0)[lt]{\lineheight{1.25}\smash{\begin{tabular}[t]{l}=\end{tabular}}}}%
    \put(0,0){\includegraphics[width=\unitlength,page=2]{imgs/unrolled_circ.pdf}}%
    \put(0.67582969,0.00571715){\color[rgb]{0,0,0}\makebox(0,0)[lt]{\lineheight{1.25}\smash{\begin{tabular}[t]{l}$(n+p)\cdot \tau$\end{tabular}}}}%
    \put(0,0){\includegraphics[width=\unitlength,page=3]{imgs/unrolled_circ.pdf}}%
    \put(0.32605143,0.01038698){\color[rgb]{0,0,0}\makebox(0,0)[lt]{\lineheight{1.25}\smash{\begin{tabular}[t]{l}$\tau$\end{tabular}}}}%
    \put(0,0){\includegraphics[width=\unitlength,page=4]{imgs/unrolled_circ.pdf}}%
    \put(-0.0022815,0.16632245){\color[rgb]{0,0,0}\makebox(0,0)[lt]{\lineheight{1.25}\smash{\begin{tabular}[t]{l}$p+n$\end{tabular}}}}%
  \end{picture}%
\endgroup%

  \caption{The sequential circuit of Fig.~\ref{fig:seq_circ} unrolled for 3 steps.
  As in Fig.~\ref{fig:seq_circ}, the first two inputs and outputs of each copy of $F$ denote the state.
  Note that the first copy of $F$ has its state inputs
    grounded to denote that $\vs_0 = (0, 0)$.}
  \label{fig:unroll}
\end{figure}

\begin{mddef}
  Let $C = (\vs_0, F)$ denote a sequential circuit with $n$ inputs, $p$ states,
  and $m$ outputs and let
  $\LAST_m : \Bool^{p + m} \to \Bool^m$ denote the bit-vector
  function that returns the last $m$ bits of input.
  For all times $\tau \in \Nat$, define the \textbf{$\tau$-unrolling}
  of $C$, to be the map:
  \begin{equation}
    \begin{split}
      &\UNROLL_C^\tau : \Bool^{\tau\cdot n} \to \Bool^m\\
      &\UNROLL_C^\tau(a_1. a_2. \ldots . a_{\tau}) \eqdef \LAST_m \circ F(\ldots F(F(F(x_0, a_1), a_2), a_3), \ldots, a_{\tau})
    \end{split}
  \end{equation}
  where each $a_i$ denotes a bit-vector in $\BoolCube$.
\end{mddef}

\begin{figure}[h]
  \centering
  \scalebox{0.65}{
    \def\svgwidth{6in}
    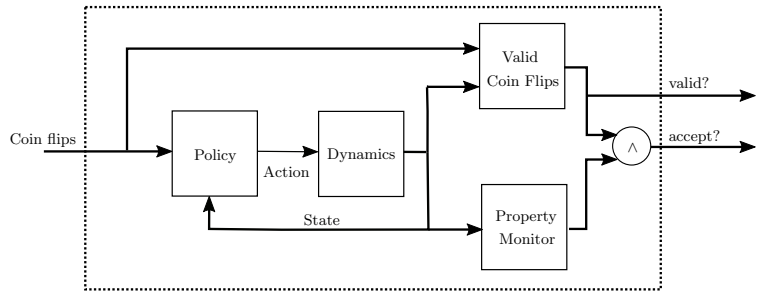
  }
  \caption{Overview of modeling dynamical system using four
sequential circuits.}
  \label{fig:monitor}
\end{figure}

Note that since unrolling a monitor results in a Boolean predicate,
Observations~\ref{obs:1},~\ref{obs:2}, and~\ref{obs:3} naturally
extend to the sequential circuits. This suggests extending our
notation for sampling a coin to sequential circuits. Namely, given a
sequential circuit $C$ and a monitor $\psi$ to condition on,
we define $\vx \tausim{\tau} C$ so that:
\begin{equation}\label{eq:seqcirc_sample}
  \Prob_{\vx \tausim{\tau} C}(\vx~|~\psi) \eqdef \Prob_{\vx \sim \UNROLL_{C}^\tau}(\vx~|~\UNROLL_{\psi}^\tau)
\end{equation}
The key utility of Eq.~\eqref{eq:seqcirc_sample}, is how it enables
studying probabilistic transition systems via model counting. In
particular, observe that queries about Finite Markov Chains
decompose into four cascading sequential circuits modeling a control
policy, a transition relation, a property monitor, and a validity
checker (see Fig.~\ref{fig:monitor}).
The dynamics circuit corresponds to the Markov Chains underlying
discrete automaton, the policy governs the transition
probabilities, the valid monitor checks that each coin flip satisfies
the conditioning property $\psi$ and the property monitor encodes
which property, $\f$, about the Markov Chain is being tested.

\begin{example}\label{ex:sailor}
  Consider the 1-d variant of the classic drunken sailor random walk.
  A sailor walks along a pier, where with each step, the sailor either
  stumbles forward by one plank, backward by one plank, or remains on
  the same plank. Further, suppose the pier is 11 planks long and
  that if the sailor visits the central plank more than 3 times, the
  plank will break and the sailor will fall into the water. If the
  sailor starts on the middle plank and the probability of moving
  forward is $2/6$, moving backward is $1/6$, and not moving is $3/6$,
  what is the probability the sailor breaks a plank after $10$ steps?

  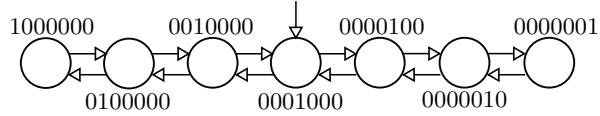
\begin{figure}
    \centering \def\svgwidth{3in} 
\begingroup%
  \makeatletter%
  \providecommand\color[2][]{%
    \errmessage{(Inkscape) Color is used for the text in Inkscape, but the package 'color.sty' is not loaded}%
    \renewcommand\color[2][]{}%
  }%
  \providecommand\transparent[1]{%
    \errmessage{(Inkscape) Transparency is used (non-zero) for the text in Inkscape, but the package 'transparent.sty' is not loaded}%
    \renewcommand\transparent[1]{}%
  }%
  \providecommand\rotatebox[2]{#2}%
  \newcommand*\fsize{\dimexpr\f@size pt\relax}%
  \newcommand*\lineheight[1]{\fontsize{\fsize}{#1\fsize}\selectfont}%
  \ifx\svgwidth\undefined%
    \setlength{\unitlength}{1068.49255659bp}%
    \ifx\svgscale\undefined%
      \relax%
    \else%
      \setlength{\unitlength}{\unitlength * \real{\svgscale}}%
    \fi%
  \else%
    \setlength{\unitlength}{\svgwidth}%
  \fi%
  \global\let\svgwidth\undefined%
  \global\let\svgscale\undefined%
  \makeatother%
  \begin{picture}(1,0.1884838)%
    \lineheight{1}%
    \setlength\tabcolsep{0pt}%
    \put(0,0){\includegraphics[width=\unitlength,page=1]{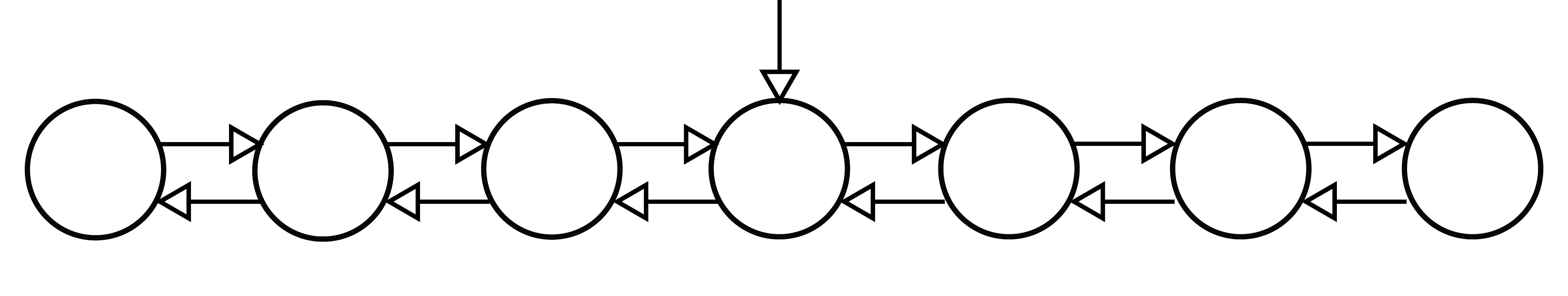}}%
    \put(0.43085486,0.00039757){\color[rgb]{0,0,0}\makebox(0,0)[lt]{\lineheight{1.25}\smash{\begin{tabular}[t]{l}0001000\end{tabular}}}}%
    \put(0.57272714,0.1313241){\color[rgb]{0,0,0}\makebox(0,0)[lt]{\lineheight{1.25}\smash{\begin{tabular}[t]{l}0000100\end{tabular}}}}%
    \put(0.27484737,0.13161849){\color[rgb]{0,0,0}\makebox(0,0)[lt]{\lineheight{1.25}\smash{\begin{tabular}[t]{l}0010000\end{tabular}}}}%
    \put(0.12956939,0.000889){\color[rgb]{0,0,0}\makebox(0,0)[lt]{\lineheight{1.25}\smash{\begin{tabular}[t]{l}0100000\end{tabular}}}}%
    \put(-0.00308462,0.13117691){\color[rgb]{0,0,0}\makebox(0,0)[lt]{\lineheight{1.25}\smash{\begin{tabular}[t]{l}1000000\end{tabular}}}}%
    \put(0.71718943,0.00443276){\color[rgb]{0,0,0}\makebox(0,0)[lt]{\lineheight{1.25}\smash{\begin{tabular}[t]{l}0000010\end{tabular}}}}%
    \put(0.87754724,0.13117691){\color[rgb]{0,0,0}\makebox(0,0)[lt]{\lineheight{1.25}\smash{\begin{tabular}[t]{l}0000001\end{tabular}}}}%
  \end{picture}%
\endgroup%

    \caption{Illustration of ``1-hot'' encoding of a chain graph. The
      right and left arrows represent arithmetic right ($\gg 1$) and
      left $(\ll 1)$ shifts of the state respectively.}
    \label{fig:chain}
  \end{figure}
  \begin{figure}
    \centering
    \scalebox{0.7}{
       \def\svgwidth{4.2in} 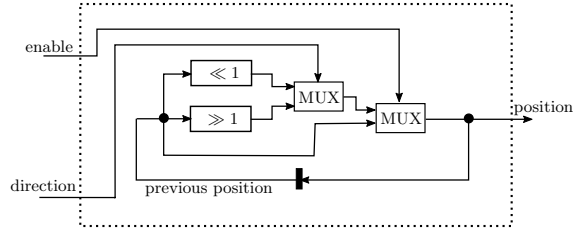
     }
    \caption{``1-hot'' encoding of a chain graph as a sequential
      circuit. The MUX gates use their top input to select which of
      their two other inputs to output.}
    \label{fig:1dgrid_circ}
  \end{figure}
  
  Within the above framework, the dynamics corresponds the
  a 1-d finite chain (see Fig.~\ref{fig:chain}). An example encoding
  of such a chain as a sequential circuit is given in Fig.~\ref{fig:1dgrid_circ}.
  Similarly, the monitor is a sequential circuit for the regular language
  $(.^*\vs_0.^*\vs_0.^*\vs_0.^*)$ which can be efficiently compiled into a sequential
  circuit~\cite{regex2circ}. Finally, the policy corresponds to some
  circuit that models the probability distribution over actions which,
  using the inputs in Fig.~\ref{fig:1dgrid_circ}), corresponds to modeling
  two independent biased coin flips, namely, $\Pr(\text{enable} = 0) = \frac{1}{2}$
  and $\Pr(\text{direction} = 0) = \frac{2}{3}$. As shown in the previous section,
  namely~\eqref{eq:arb_coin}, we can model the direction coin by feeding the output of
  $\f_{\text{direction}}(\vx) = \vx < 2$ into the direction input shown in Fig.~\ref{fig:1dgrid_circ},
  and conditioning on $\vx < 3$, where $\vx \in \Bool^2$. Similarly, using a disjoint set
  of inputs, on can encode the enable coin by feeding the output of $\f_{\text{enable}}(\vx') = \vx' < 1$
  into the enable input of Fig.~\ref{fig:1dgrid_circ}. The resulting sequential circuit, $C_{ds}$ is summarized
  in Fig.~\ref{fig:drunken_sailor}.
  \begin{figure}
    \centering
      \scalebox{0.7}{
        \def\svgwidth{\linewidth} 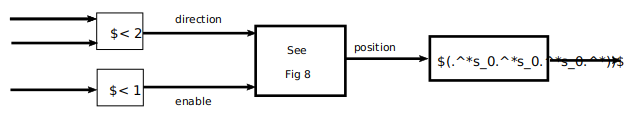
      }
    \caption{Drunken Sailor policy circuit, $C_{ds}$.}
    \label{fig:drunken_sailor}
  \end{figure}
  Letting $C_{valid}$ denote the monitor that for all time steps, $\vx
< 3$, then the probability of the sailor falling into the water within
the first 50 steps is given via:
  \begin{equation}
    \Prob\;(\text{sailor falls into water}) = \Prob_{~~~\mathclap{x \tausim{50} C_{ds}}}\;(x = 1~|~C_{valid}) = \frac{25398396}{6^{10}} \approx 0.42
  \end{equation}
  Finally, it is worth noting that anecdotally, this problem took
  $\approx$ 1 minute using a BDD and $\approx$ 1 sec using the  SAT
  based approximate model counter ApproxMC3~\cite{SM19}.
\end{example}


\section{Distributions over Finite Sets}\label{sec:arb_dist}
We now return to the topic of modeling distributions over finite sets
by feeding coin flips into circuits. The first two techniques
can be used to model arbitrary rational-valued distributions over
finite sets. Then, for variety, we illustrate how to encode
a Binomial distribution as a sequential circuit.

Formally, we first seek to systematically solve the following problem:
\begin{mdframed}
  \begin{problem}\label{prob:encode}
    Let $Y$ be a finite set whose elements, $\hat{y}_i$, are numbered
    from $1$ to $|Y|$, and associate to $Y$ the following rational
    valued probability distribution:
    \begin{equation}
      \Prob(\hat{y}_i) = \nicefrac{a_i}{m},
    \end{equation}
    where $a_i, m \in \Nat$ such that
    $\displaystyle \sum_{i=1}^{|Y|} a_i = m > 0$.  Further, denote by
    $\vy \in \Bool^{|Y|}$ the $1$-hot encoding of elements of $Y$,
    e.g.  $y_i = 1$ iff $\vy$ corresponds to $\hat{y}_i$. Find an
    $n\in \Nat$, an $n$-bit-vector function
    $F : \BoolCube \to \Bool^{|Y|}$, and a $n$-bit-vector predicate $\psi$,
    such that:
    \begin{equation}
      \Prob(\hat{y}_i) = \Prob_{\mathclap{\vy \sim F}}\;(y_i = 1~|~\psi)
    \end{equation}
  \end{problem}
\end{mdframed}
Note that the use of a $1$-hot encoding is without loss of generality,
since one can always feed this encoding into a circuit that transforms
it into another encoding.

\mypara{Common Denominator Method}
Our first technique is a straightforward generalization of encoding a
biased coin~\eqref{eq:arb_coin}. The key idea is to encode $|Y|$
mutually exclusive biased coins, which together, form
a $1$-hot encoding of $\hat{y}_i$.
To begin, let $n$ be the smallest integer such that $m \leq 2^n$.
For convenience, define
\begin{equation}
  b_0 \eqdef 0~~~~~b_{i+1} \eqdef b_{i} + a_i.
\end{equation}
Now, let
$\f_i : \Bool^n \to \Bool$ denote the circuit,
\begin{equation} \label{eq:mux_gadget}
  \f_i(\vx) \eqdef b_i \leq \vx < b_{i} + a_i
\end{equation}
where $\vx$ is interpreted as an unsigned integer. Further, note that
by construction, $\#(\f_i) = a_i$ and the $\f_i$ are mutually
exclusive.
Thus, the product of all $\f_i$ results in a $1$-hot encoding.
Namely, letting $F : \Bool^n \to \Bool^{|Y|}$ denote $\f_1 \times \ldots \times \f_{|Y|}$
and $\psi(\vx) \eqdef \vx < m$ yields,
\begin{equation}
  \Prob_{\mathclap{\vy \sim F}}\;(y_i = 1~|~\psi) = \nicefrac{a_i}{m},
\end{equation}
as desired. Finally, before discussing our second technique, we
briefly remark that it was using this technique the encodings seen so
far have been generated. For example, the circuit seen in
Ex.~\ref{ex:mux} is a straightforward simplification of the circuit
formed by feeding the circuit created using, $a_0 = 3$, $a_1 = 1$,
$a_2 = 4$, $m = 2^3$, and $n=3$ into a circuit which transforms the
$1$-hot encoding of unsigned integers into the base 2 encoding.

\mypara{Knuth and Yao Random Number Generators} As the reader may be
aware, simulating arbitrary discrete distributions using coin flips is
well-trodden ground. For example, Knuth and Yao famously provided a
systematic technique for simulating arbitrary discrete distributions
in a manner that (in expectation) is optimal with respect the number
of coin flips required~\cite{knuthyaodice}. As an example of the
flexibility of the above framework, we shall sketch how to embed the
Knuth and Yao's scheme as a sequential circuit with coin flip inputs.
\begin{wrapfigure}{r}{0.35\textwidth}
  \centering
  \scalebox{1}{
  \def\svgwidth{1.3in}
  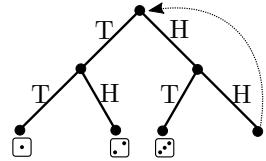
  }
  \caption{Illustration of infinite parse tree of a fair three sided
    die. Note that back edges represent self similar subtrees.}
  \label{fig:parse_tree}
\end{wrapfigure}

As before, we assume the set-up given in Problem~\ref{prob:encode} and
define $n$ to be the smallest integer such that $m \leq 2^n$.
Next, write each probability mass in its binary expansion,
\begin{equation}
  \Prob(\hat{y}_i) = 0.p_1p_2p_3\ldots.
\end{equation}
For example, if $\Prob(\hat{y}_i) = 2^{-3}$, then the correspond
decimal expansion is $0.001$.
Similarly, $\Prob(\hat{y}_i) = 1/3$
yields $0.(01)^\omega$, where $(\cdot)^\omega$ represents an infinite
repetition. Knuth and Yao's key idea is to then construct a
(potentially infinite) binary tree where if the $j$th bit of the
expansion of $\Prob(\hat{y}_i)$ is 1, then $\hat{y}_i$ appears as a
leaf at depth $j$.  Such a tree is guaranteed to exist due to the
Kraft inequality~\cite{kraft1949device}.

For example, for a three sided die, $\Prob(\epsdice{1}) =
\Prob(\epsdice{2}) = \Prob(\epsdice{3}) = 0.(01)^\omega$ with the
corresponding of the infinite binary tree shown in
Fig.~\ref{fig:parse_tree}. Note that if a sub-tree is self similar to
an ancestor node, we draw a back edge. We refer to the tree with back
edges as a \textbf{parse tree}.  Knuth and Yao's algorithm then performs
a depth first search from the root to a leaf where at each node, one
flips a coin and takes the left branch if the coin comes up tails
and take the right branch otherwise. Once a leaf is reached, the algorithm
then outputs the leaf's value.

The central idea in porting this algorithm to our framework is to
encode the transition system given by the depth first search on the
parse tree into a sequential circuit. This is done by viewing the
parse tree as $|Y|$ monitors, $C_i$, each accepting iff
the corresponding element has been reached. For example, for the three sided
dice, the parse tree given in Fig.~\ref{fig:parse_tree} results in
three monitors corresponding to recognizing $\epsdice{1}, \epsdice{2}$,
and, $\epsdice{3}$ respectively. These $|Y|$ monitors are then fed the
same stream of random coin flips and have their outputs concatenated
to for the $1$-hot encoding of $Y$; however, note that until a leaf
is reached, the resulting circuit, $C$, will output the all zeros
bit-vector, $0^{|Y|}$.

Finally, to create a model counting problem, one observes that
(asymptotically) the probability of not having reached a leaf state
exponentially decreases with the number of coins flipped. Thus, if
$\tau \in \Nat$ is sufficiently large and $\psi$ denotes the monitor
checking if the last state is $0^{|Y|}$ then,
\begin{equation}
  \Prob(\hat{y}_i) \approx \Prob_{\vy \tausim{\tau} C}(y_i=1~|~\psi).
\end{equation}
While the utility of this method may seem suspect, we note that (i.)
For many cases, $\tau$ simply needs to be the height of the tree.  For
example, if $m$ is a power of $2$, then no back edges will exist.
Similarly, if all back edges occur at leaves and go to the root, as in
Fig.~\ref{fig:parse_tree}, then the back edges can be safely removed by
conditioning on $\psi$. In fact, many times, such as Fig.~\ref{fig:parse_tree},
this encoding is equivalent to technique 1!
(ii.) The parse tree automatically
takes into account the particularities of the distribution and as
previously stated, is known to be optimal in expectation, which
translates to optimality when no back edges are present. (iii.)
Finally, and most importantly, this example illustrates how the
literature on transforming discrete distributions could shape
the design of model counting encodings of probabilistic systems.
With this connection to prior literature explored, we now
evaluate how the framework developed relates to prior work on
reducing weighted model counting to unweighted model counting.

\mypara{Binomial Distribution}
For our final technique, we illustrate how to encode a Binomial
Distribution as a probabilistic circuit. Formally,
let $X$ be the number of successes after $n$ independent trials,
the Binomial Distribution with bias $p$ is defined by:
\begin{equation}\label{eq:binom}
  \Prob(X = k) = {{n}\choose{k}}p(1 - p).
\end{equation}

\begin{wrapfigure}{r}{0.35\textwidth}
  \centering
  \scalebox{1}{
  \def\svgwidth{1.3in}
  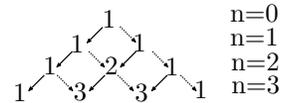
  }
  \caption{Visualization transition relation modeled by Fig.~\ref{fig:binom} circuit for n=3. Left arrows represent trial
    failures, right arrows represent trial successes.}
  \label{fig:pascal}
\end{wrapfigure}
\begin{figure}[h]
  \centering
  \scalebox{0.8}{
    \def\svgwidth{3.5in}
\begingroup%
  \makeatletter%
  \providecommand\color[2][]{%
    \errmessage{(Inkscape) Color is used for the text in Inkscape, but the package 'color.sty' is not loaded}%
    \renewcommand\color[2][]{}%
  }%
  \providecommand\transparent[1]{%
    \errmessage{(Inkscape) Transparency is used (non-zero) for the text in Inkscape, but the package 'transparent.sty' is not loaded}%
    \renewcommand\transparent[1]{}%
  }%
  \providecommand\rotatebox[2]{#2}%
  \newcommand*\fsize{\dimexpr\f@size pt\relax}%
  \newcommand*\lineheight[1]{\fontsize{\fsize}{#1\fsize}\selectfont}%
  \ifx\svgwidth\undefined%
    \setlength{\unitlength}{887.98484982bp}%
    \ifx\svgscale\undefined%
      \relax%
    \else%
      \setlength{\unitlength}{\unitlength * \real{\svgscale}}%
    \fi%
  \else%
    \setlength{\unitlength}{\svgwidth}%
  \fi%
  \global\let\svgwidth\undefined%
  \global\let\svgscale\undefined%
  \makeatother%
  \begin{picture}(1,0.27394523)%
    \lineheight{1}%
    \setlength\tabcolsep{0pt}%
    \put(0.3630486,0.19323803){\color[rgb]{0,0,0}\makebox(0,0)[lt]{\lineheight{1.25}\smash{\begin{tabular}[t]{l}$\ll 1$\end{tabular}}}}%
    \put(0.3215341,0.25970064){\color[rgb]{0,0,0}\makebox(0,0)[lt]{\begin{minipage}{0.16978359\unitlength}\raggedright \end{minipage}}}%
    \put(0,0){\includegraphics[width=\unitlength,page=1]{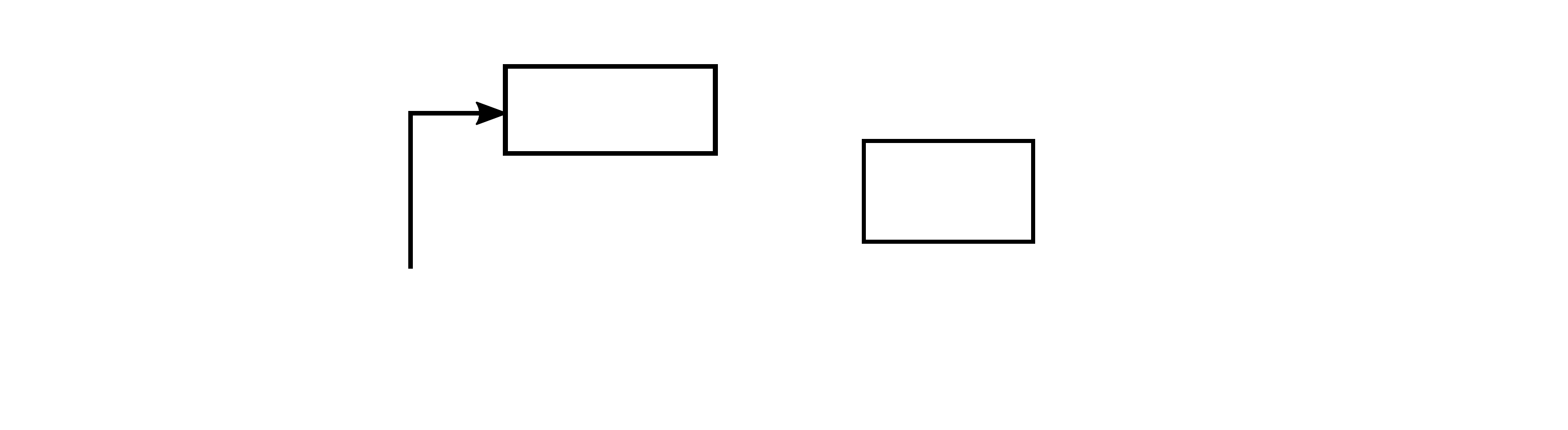}}%
    \put(0.56052814,0.13869982){\color[rgb]{0,0,0}\makebox(0,0)[lt]{\lineheight{1.25}\smash{\begin{tabular}[t]{l}MUX\end{tabular}}}}%
    \put(0,0){\includegraphics[width=\unitlength,page=2]{imgs/binom.pdf}}%
    \put(0.36392167,0.02606919){\color[rgb]{0,0,0}\makebox(0,0)[lt]{\lineheight{1.25}\smash{\begin{tabular}[t]{l}previous 1-hot count\end{tabular}}}}%
    \put(-0.0009073,0.15349191){\color[rgb]{0,0,0}\makebox(0,0)[lt]{\lineheight{1.25}\smash{\begin{tabular}[t]{l}trial\end{tabular}}}}%
    \put(0,0){\includegraphics[width=\unitlength,page=3]{imgs/binom.pdf}}%
    \put(0.80573995,0.16377275){\color[rgb]{0,0,0}\makebox(0,0)[lt]{\lineheight{1.25}\smash{\begin{tabular}[t]{l}1-hot count\end{tabular}}}}%
  \end{picture}%
\endgroup%

  }
  \caption{Sequential Circuit for counting trials. The latch
  is initialized with $1$ and every time a successful trial occurs (trial = 1) the state is left shifted. For example, if $n=3$, then $\vec{s}_0 = 001$. If a successful trial occurs then $\vec{s}_1 = 010$.}
  \label{fig:binom}
\end{figure}

Aside from being an interesting example, Binomial distributions are an
important building block for many probabilistic systems. Of particular
interest is the common use of~\eqref{eq:binom} to approximate Gaussian
distributions when discretizing continuous domains.
The key idea in encoding~\eqref{eq:binom} is to feed biased coins into
a circuit that counts the number of successful trials (out of $n$). An
example encoding is given in
Fig.~\ref{fig:binom}. Intuitively, this circuit can be
visualized as a modeling the transitions of pascals triangle (Fig.~\ref{fig:pascal}).  Given
such a sequential circuit, $C$, the encoding for a $p=\nicefrac{1}{2}$
Binomial Distribution of $n$ trials is simply the $\tau=n$ unrolling
of $C$. Therefore,
if $\f_k(\vx) \eqdef (\vx = k)$  and $F_{n}(\vx) \eqdef \UNROLL_C^n(\vx)$ then:
\begin{equation}
  \Prob(X = k) = \Prob_{y \sim \f_k \circ F_n}(y = 1) = \frac{\#(f_k \circ F_n)}{2^n}.
\end{equation}
Finally, observe that by feeding the output of the
circuit encoding of a biased coin~\eqref{eq:arb_coin} into
the trial input of $C$ enables encoding any rational $p$ Binomial Distribution through appropriate unrolling.


\section{Relationship to Weighted Model Counting}\label{sec:rel_unweighted}

In the preceding sections, we have developed a modeling framework for probabilistic systems based on feeding \emph{unbiased} coins into Boolean predicates or sequential circuits.
Our encodings require only \emph{unweighted} model counting algorithms for their analysis, and thus directly benefit from the recent dramatic performance gains in unweighted model counting~\cite{chakraborty2016algorithmic,achlioptas2018fast,SM19}.
In contrast, many previous works on probabilistic inference using model counting algorithms have built on algorithms for \emph{weighted model counting}.
But adapting advances in unweighted model counting to weighted model counting can be quite challenging, and we are not aware of clear performance benefits that can be gained by considering weighted model counting in the algorithm itself.
Instead, it appears to be easier to reduce weighted model counting to unweighted model counting~\cite{weighted2unweighted}.

In this section, we take a second look at previously proposed encoding of weighted model counting in unweighted model counting.
Here, we focus on literal-weighted model counting, which extends unweighted model counting by a function $W$ mapping each pair of input coin and Boolean outcome to a real value.
The \emph{weight} of a model is then defined as the product of the weights of its components.
For example, consider the weight function assigning $W(x_1)=0.2$, $W(\overline x_1)=1$, $W(x_2)=3$, and $W(\overline x_2)=15$ the model $x_1\wedge \overline x_2$ has thus weight $W(x_1)\cdot W(\overline x_2)=3$.

Chavira and Darwiche have shown that this general setting can be efficiently reduced to the case that $W(x)\in[0,1]$ and $W(x)=1-W(\overline{x})$~\cite{chavira2008probabilistic}.
In particular, this means that we can simply model weighted literals by biased coins (Section~\ref{sec:rand_inputs}).

The curious reader may ask how this compares to the encodings proposed in prior work~\cite{weighted2unweighted}.
There, the authors build on the following gadget:
%
%
%
%
%
\begin{mddef}
  Let $k_n k_{n-1} \ldots k_1$ be the standard base 2 representation of $k$ as an unsigned integer with $k_1$ being the least significant bit. 
  We define
\begin{equation}
  \KMODELS^k_n(\vx) \eqdef x_n \square_n \KMODELS^k_{n-1}(\vx),
\end{equation}
where $\KMODELS^k_0(\vx) = 0$, $\square_n \eqdef \wedge$ if $k_n= 0$ and $\vee$ otherwise.
\end{mddef}
The key utility of $\KMODELS^k_n$ is that $\#(\KMODELS^k_n) = k$~(see~\cite{weighted2unweighted}).
The reduction from weighted to
unweighted model counting then straightforwardly follows from,
\begin{equation}
  \#\big(\KMODELS^k_n(\vy) \wedge \f(\vx)\big) = k\cdot \#\f(\vx).
\end{equation}
In particular, if $\f(\vx)$ is given as a CNF formula, then applying
the above reduction to each clause is structurally equivalent to the
reduction given in~\cite{weighted2unweighted}.

We simply want to observe here that $\KMODELS^k_n(\overline{\vx})$ satisfies the same set of models as the bit-vector comparison $\vec x < k$.
\begin{mdframed}
    \begin{proposition}
      \begin{equation}\label{eq:lemma_kmodels}
        \KMODELS_n^k(\overline{\vx}) \equiv \vx < k,
      \end{equation}
      where $\overline{\vx}$ indicates the bit-wise negation of $\vx$.
    \end{proposition}
\end{mdframed}
\begin{proof}
  Observe that one can test if $\vx < k$ by recursively testing if
  the most significant bit of $\vx$ is less than the most significant
  bit of $k$ in base 2 representation, i.e,
  \begin{equation}\label{eq:lt_1}
    \widehat{\KMODELS}_n^k(\vx) \eqdef (x_n < k_n) \vee \Big(x_n = k_n \wedge \widehat{\KMODELS}_{n-1}^k(\vx)\Big)
  \end{equation}
  where $\widehat{\KMODELS}_x^0(0) \eqdef 0$ and
  $(x_n < k_n) \equiv (\neg x_n \wedge k_n)$.
  Observe that if $k_n = 0$, then that~\eqref{eq:lt_1} reduces to,
  $\widehat{\KMODELS}_n^k(\vx) =  \neg x_n \wedge \widehat{\KMODELS}_{n-1}^k(\vx)$.
  Similarly, if $k_n = 1$ then,
  \begin{equation}
      \widehat{\KMODELS}_n^k(\vx) = \neg x_n \vee \Big(x_n  \wedge \widehat{\KMODELS}_{n-1}^k(\vx)\Big)
      = \neg x_n \vee \widehat{\KMODELS}_{n-1}^k(\vx).
  \end{equation}
  Defining, $\square_n \eqdef \vee$ if $k_n = 1$ and $\square_n \eqdef \wedge$ if $k_n = 0$, then~\eqref{eq:lt_1} becomes,
  $\widehat{\KMODELS}_n^k(\vx) = \neg x_n~{\square}_n~\widehat{\KMODELS}_{n-1}^k(\vx)$, which is equivalent to $\KMODELS_n(\overline{\vx})$. \qed
\end{proof}
\noindent

In particular, this shows that the basis of this work, the encoding of biased coins via unbiased coins, is equivalent to prior encodings.

\section{Conclusion}

In this paper, we systematically developed a framework for modeling
probabilistic systems as model counting problems.  Starting from
unbiased coins, we construct biased coins, correlated coins, and
conditional probabilities.  We discuss how to model arbitrary
distributions over finite sets and how to combine our building blocks
into sequential systems.  While the building blocks discussed in this
work have been used in previous works
(e.g.~\cite{vc,rabe2014symbolic}), we believe that the explicit
discussion of the modeling techniques in this work will enable future
case studies on probabilistic systems with SAT-based model counting
algorithms. For example, as illustrated by the Knuth Yao and binomial
encodings, probabilistic sequential circuits an excellent mechanism to
adapt techniques from the larger random bit model literature into
\#SAT encodings. Finally, implementations for the common denominator
method and the binomial distribution method can by found at \url{https://github.com/mvcisback/py-aiger-coins} and are implemented
using the py-aiger library~\cite{pyAiger}.


%
%
%
%
\bibliographystyle{splncs04}
\bibliography{refs}

\end{document}